\documentclass[
    ,final            
  ]
  {aipproc}

\layoutstyle{6x9}
\usepackage{graphicx}
\usepackage{epstopdf}
   \usepackage{color}
\def\apj{ApJ}
\def\apjl{ApJL}
\def\apjsupp{ApJ Supp.}

\def\mnras{{M.N.R.A.S.}}

\newcommand{\vol}[2]{$\,$\rm #1\rm , #2.}           

\begin{document}
\bibliographystyle{plainnat}

\title{Population Synthesis of Radio and Gamma-ray Pulsars in the Fermi Era}



\author{Peter L. Gonthier}{
  address={Hope College, Department of Physics, Holland, MI, USA}
}

\author{Caleb Billman}{
  address={Hope College, Department of Physics, Holland, MI, USA}
}

\author{Alice K. Harding}{
  address={NASA Goddard Space Flight Center, Greenbelt, MD, USA}
}

\author{Isabelle A. Grenier}{
  address={AIM, Service dÕAstrophysique, CEA Saclay, France}
}
\author{Marco Pierbattista}{
  address={AIM, Service dÕAstrophysique, CEA Saclay, France}
}

\begin{abstract}
 We present results of our pulsar population synthesis of normal pulsars from the Galactic disk using our previously developed computer code.  On the same footing, we use slot gap and outer gap models for gamma-ray emission from normal pulsars to obtain statistics of radio-loud and radio-quiet gamma-ray pulsars.   From recently improved understanding of HII and star forming regions in the Galaxy, we develop a new surface density model of the birth location of neutron stars.   We explore models of neutron star evolution with magnetic field-decay, and with different initial period and magnetic field distributions.  We present preliminary results including simulated population statistics that are compared with recent detections by Fermi of normal, isolated pulsars.
\end{abstract}

\maketitle


\section{INTRODUCTION}

 NASA's new flagship, the Fermi Gamma-Ray Space
Telescope ({\it Fermi}) launched in June 2008, has opened a new era in $\gamma$-ray astronomy,  improving dramatically our understanding of $\gamma$-ray emission from pulsars.  {\it Fermi} has discovered over 70 normal, isolated and millisecond pulsars, many through blind period searches,  superseding the $\gamma$-ray pulsar database of six provided by its predecessor, the EGRET instrument aboard the Compton Gamma-Ray Observatory.   A group of 40 normal, isolated pulsars that are  publicly available are used in this study.

\section{Birth Distributions}

Previous to this study, we have used \citep{Gont} the exponential, spatial distributions from the work of Paczy\'nski \citep{Pac}.  Our preliminary statistics of $\gamma$-ray pulsars indicated that our simulations required a larger group of nearby pulsars, as the code was not simulating sufficient {\it Fermi} pulsars.  As a result, we developed a new surface density distribution of neutron star birth locations adopting the recent discoveries of nearby H II regions \citep{Bania}.  We display in Figure 1 a comparison of the smooth exponential probability radial distribution  \citep{Pac} represented by a dashed curve, and our new model shown as a solid curve.  This new radial distribution together with the exponential height distribution and a uniform angular distribution are used to generate randomly distributed neutron stars in the Galaxy at birth.  We implement the supernova kick velocity distribution by Hobbs et al. \citep{Hobbs}, and evolve the neutron stars in the Galactic potential \citep{Pac} using a a Cash-Karp fifth order Runge-Kutta routine \citep{Press}.   The birth rate of ~2 neutron stars per century \citep{Tamm} back to 1 Gyr in the past is assumed to be constant.

\begin{figure}
  \includegraphics[height=.25\textheight]{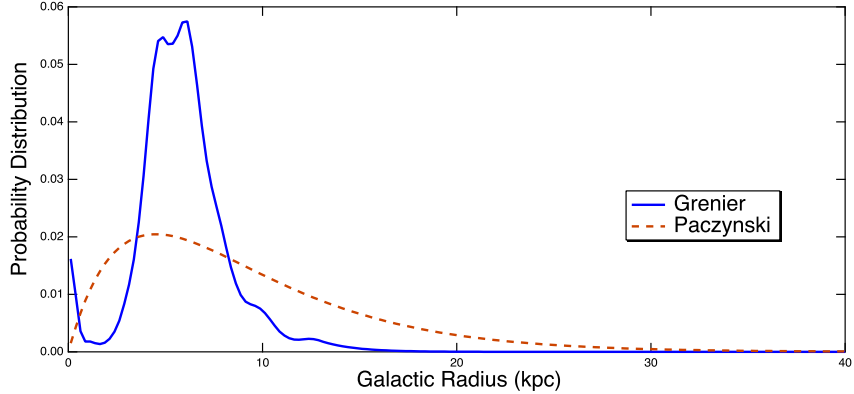}
  \caption{Radial birth probability as a function of the Galactic radius in kpc (cylindrical coordinates).  The dashed curve represents
  the probability distribution given by the study of Paczy\'nski \citep{Pac}, while the solid curve represents our new model probability distribution   implementing the newly discovered HII regions in the study of Bania et al. \citep{Bania}.}
\end{figure}

We assume a Gaussian distribution for the initial period distribution of the neutron stars with a mean $\hat P$ of 50 ms and a width $\sigma$ of 50 ms in the form:
\begin{equation}
P(P_o) \propto e^{ - (\hat P   - P_o)^2 /\sigma^2 } 
\end{equation}
The initial magnetic field distribution is represented by a sum of two log-normal distributions as
\begin{equation}
P(B) \propto \sum\limits_{i = 1}^2 {A_i e^{ - (\mu _i  - \log (B))^2 /\sigma _i^2 } } 
\end{equation}
with the parameters given in Table 1.  In order to improve the agreement between the simulated pulsars and those detected by {\it Fermi} in the $\dot P- P$ diagram, we had to change the widths and means  the initial magnetic field and period distributions from our previous work in order to produce a greater number of younger, high $\dot E$ pulsars.  We assume a field decay constant of 2.8 Myr that is similar to our previous studies.  We are not necessarily implying the existence of field decay, rather we suggest that field decay may be mimicking some form of pulsar spin-down.
\begin{table}
\begin{tabular}{lcccccc}
\hline
Initial Magnetic Field  & \tablehead{1}{c}{b}{$A_1$}
  & \tablehead{1}{c}{b}{$\mu_1$}
  & \tablehead{1}{c}{b}{$\sigma_1$}
   & \tablehead{1}{c}{b}{$A_2$}
 & \tablehead{1}{c}{b}{$\mu_2$}
  & \tablehead{1}{c}{b}{$\sigma_2$} \\
\hline
Previous Study \cite{Gont} & .6 & 12.6 & 0.65 & 0.3 & 13.0 & 0.8 \\
Current Study & .6 & 12.5 & 0.4  & 0.35 & 13.1 & 0.4 \\
\hline\hline
 Initial Period && \tablehead{1}{c}{b}{$\hat P$} 
  && \tablehead{1}{c}{b}{$\sigma$}   \\
\hline
Previous Study \cite{Gont} && 300 ms && 300 ms \\
Current Study && 50 ms && 50 ms \\
\hline
\end{tabular}
\caption{Assumed initial parameters of previous \cite{Gont} and current studies for the birth magnetic field (log Normal) and period (Gaussian) distributions.}
\label{tab:a}
\end{table}

\section{Simulation and Results}

In the process of improving the agreement with the {\it Fermi} detected pulsars, we find that the comparisons with the radio pulsar distributions have had to be compromised to some extent as can be seen in Figure 2.  We represent the distributions of the indicated characteristics of the detected radio pulsars in solid histograms while the distributions of simulated pulsars are indicated in open histograms. The new radial birth distribution shown in Figure 1 leads to an excess of radio pulsars between the Sun and the Galactic center as shown in the comparisons of the distance histograms in Figure 2.  We also simulate an excess of young radio pulsars as indicated in the period, period derivative and age histograms.
\begin{figure}
  \includegraphics[height=.35\textheight]{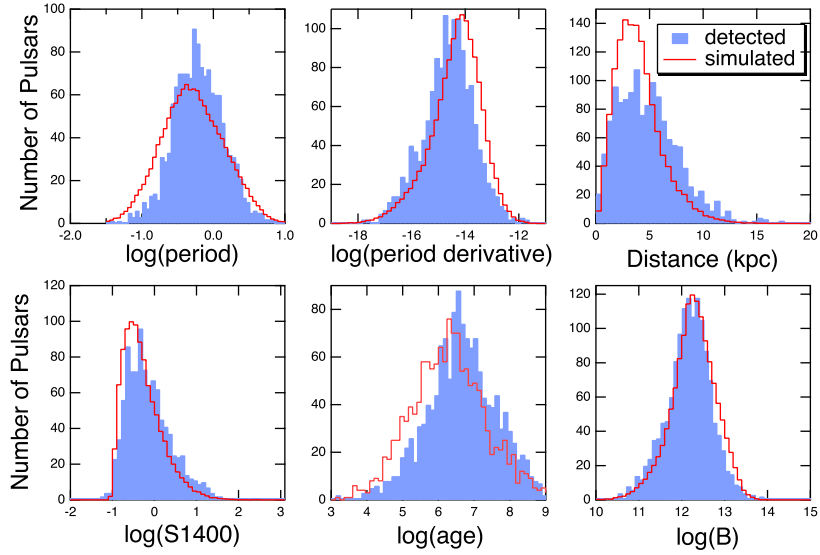}
  \caption{Histograms comparing the indicated pulsar characteristics for the group of  detected (solid) and the group of simulated (open) radio pulsars.}
\end{figure}

In Figure 3 (left panel), we present the detected radio pulsars (solid dots) observed from a select group of ten radio surveys, and the detected {\it Fermi} radio-loud (bowties) and radio-weak (hourglasses), where the definition of radio-weak is used to indicate that these pulsars were not detected in our select group of radio surveys.  In the right panel, we show the simulated radio and $\gamma$-ray pulsars.  The $\gamma$-ray emission used here is from the polar cap/Slot Gap (SG) model.  We normalize the simulation to the number of detected radio pulsars observed in the ten radio surveys.  The radio luminosity is adjusted to produce a neutron star birth rate of ~2.1 per century \cite{Tamm}.  Further details of the $\gamma$-ray beam geometry and luminosity as well as alternate $\gamma$-ray models are presented in the companion paper of this conference proceedings by Pierbattista et al.  Here we are using a group of 30 radio-weak and 10 radio-loud {\it Fermi} pulsars that have been made public.  Our simulation predicts a slightly smaller ratio of radio-weak to radio-loud with 44 radio-loud and 17 radio-weak $\gamma$-ray pulsars.  Preliminary studies suggest that it should be possible to simulate both {\it Fermi} and radio pulsars with the same group of evolved neutron stars.  However, further refinements are required in order to improve the agreement between detected and simulated distributions.

\begin{figure}
  \includegraphics[height=.33\textheight]{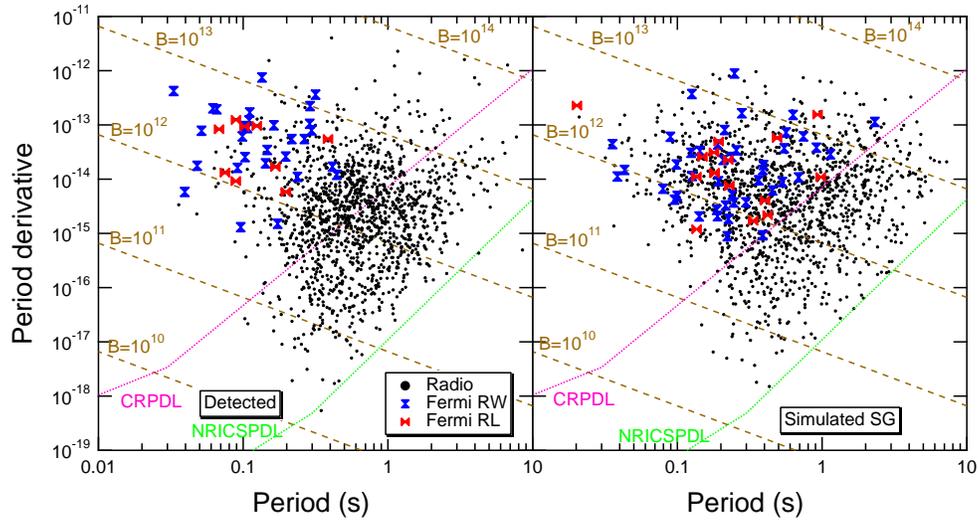}
  \caption{Period derivative - period plot of detected (left) and simulated (right) pulsars.  Radio pulsars indicated in solid dots, {\it Fermi} radio-loud (bowties) and radio-weak (hourglasses) pulsars.  The dashed lines represent constant magnetic field loci assuming vacuum, dipole spin-down, and dotted lines indicate curvature radiation and non-resonant Compton Scattering pair death lines.}
\end{figure}

\begin{theacknowledgments}
 We express our gratitude for the generous support of the Michigan Space Grant Consortium, of the National Science Foundation (REU Grant No. PHY/DMR-1004811), the NASA Astrophysics Theory and Fundamental Program (NNX09AQ71G) and the NASA Fermi Guest Investigator Cycle 3 Program (NNX10AO41G).
 \end{theacknowledgments}



\begin{thebibliography}{9}   


\bibitem[Bania et al. (2010)]{Bania}
Bania, T.M. et al. 2010, \apjl, \vol{718}{L106}

\bibitem[Gonthier et al. (2009)]{Gont}
Gonthier, P. L. et al. 2009, {\it 2009 Fermi Symposium eConf Proceedings C091122}, arXiv:0912.3539

\bibitem[Hobbs et al. (2005)]{Hobbs}
Hobbs, G., Lorimer, D. R., Lyne, A. G., \& Kramer, M. 2005, \mnras, \vol{360}{974}

\bibitem[Tammann et al. (1994)]{Tamm}
Tammann, G.A., L\"offler, W., \& Schr\"oder, A. 1994, \apjsupp, \vol{92}{487}

\bibitem[Paczy\'nski (1990)]{Pac}
Paczy\'nski, B. 1990, \apj, \vol{348}{485}

\bibitem[Press et al. (1992)]{Press}
Press, W. H. et al. 1992, {\it Numerical Recipes in C The Art of Scientific Computing}, (Cambridge
University Press New York), 714

\end{thebibliography}
\end{document}